\documentstyle[twocolumn,twoside,epsf]{IEEEtran}  
\begin{document}               
\title{The Channel Capacity of a Fiber Optics Communication System: 
perturbation theory}
\author{Evgenii Narimanov and Partha Mitra}
\maketitle


\begin{abstract}   
We consider the communication channel given by a 
fiber optical transmission line. We develop a method to perturbatively
calculate the information capacity of a nonlinear channel, given the
corresponding evolution equation. Using this technique, we 
compute the decrease of the channel capacity to the leading order in
the perturbative parameter for fiber optics communication systems.
\end{abstract}

\vskip 0.5 truecm

\section{Introduction}

\PARstart{T}he 
performance of any communication system is ultimately limited by the
signal to noise ratio of the received signal and available bandwidth. This
limitation can be stated more formally by using the concept of {\it channel
capacity} introduced withing the framework of information theory\cite{Shannon}.
The channel capacity is defined as the maximum possible bit rate for
error-free transmission
 in the presence of noise. For a linear communication
channel with additive white Gaussian noise, and a total signal power
constraint at the input, the capacity is given by the celebrated 
Shannon formula\cite{Shannon}
\begin{eqnarray}
C & = & W \log\left(1 + \frac{P_0}{P_N}\right)
\label{eq:Shannon}
\end{eqnarray}
where $W$ is the channel bandwidth, $P_0$ is the average signal power, 
and $P_N$ is the average noise power.

Current optical fiber systems operate substantially below the fundamental
limitation, imposed by the Eq. (\ref{eq:Shannon}). However, a 
considerable improvement in the coding schemes for lightwave communications,
expected in the near future, may result in the development of systems, whose
efficiency may approach this fundamental limit. 

However, the representation of the channel capacity in the standard 
form (\ref{eq:Shannon}) 
is unsuitable for applications to the actual fiber optics systems.
It  was obtained based on the assumption of {\it linearity} of the 
communication channel, while the modern fiber optics systems operate in
a substantially nonlinear regime. Since the optical transmission lines
must satisfy very strict requirements for bit-error-rate ($10^{-12}$ to 
$10^{-15}$), the pulse amplitude should be large enough so that is can be 
effectively detectable. The increase of the number of wavelength-division 
multiplexing (WDM) channels\cite{Agrawal} in the modern fiber optics
communication systems also leads to a substantial increase of the electric
field intensity in the fiber. As a consequence, the Kerr nonlinearity of 
the fiber 
refractive index $n = n_0 + \gamma I$ (where $I$ is the pulse intensity) 
becomes substantial and should be taken into account.

In the present paper we consider 
corrections to the channel capacity of the optical fiber communication system,
originating from the nonlinearity of the fiber. The technique that we use
involves a perturbative computation of the relevant mutual information and
subsequent optimization. To our knowledge, this method appears to be
substantially new.

\section{Fiber Optics Communication System as an Information Channel}

We consider a typical fiber optics communication system, which consists 
of a sequence of $N$ fibers each 
followed by an amplifier (see Fig. 1). 
The amplifiers have to be introduced in order to 
compensate for the power loss in the fiber. An inevitable consequence of 
such design, however, is the generation of the noise in the system, 
coming from the spontaneous emission in the optical amplifiers.
For simplicity, we will assume that all the fibers and the amplifiers 
of the link are identical.

The information is encoded in the electric field at the ``imput'' of the 
system, typically using the light pulses sent at different frequencies.
The available bandwidth of the amplifiers as well as the increase of the fiber
absorption away from the ``transparency window'' near the wavelength 
$\lambda = 1.55 \mu m$, limits the bandwidth of the 
fiber optic communication system.

The maximum amount of the information, that can be
 transmitted through the communication
system per unit time, is called the channel capacity $C$. 
According to the Shannon's basic result\cite{Shannon}, this quantity
is given by the maximum value of the 
mutual information per second over all possible
input distributions:
\begin{eqnarray}
C & = &  {\rm max}_{p_x} 
\left\{ H\left[{y}\right] - 
\langle H\left[{y}\left|\right.{x}\right] 
\rangle_{p_x} \right\}
\end{eqnarray}
The mutual information 
\begin{eqnarray}
R & = &  H\left[{ y\left(\omega\right)}\right] - 
\langle H\left[{y\left(\omega\right)}|{x\left(\omega\right)}\right] 
\rangle_{p_x}
\end{eqnarray}
is a functional of the ``input distribution'' 
$p_x\left[x\left(\omega\right)\right]$, which represents the encoding 
of the information using the electric field components at different 
frequences
\begin{eqnarray}
E_{\rm in}\left(t\right) & = & \int_W d\omega \ x\left(\omega\right) 
\exp\left(i \omega t\right) 
\end{eqnarray}
The entropy $H\left[{y\left(\omega\right) }\right]$ is 
the measure of the information
received at the output of the communication channel. However, if the 
channel is noisy, for any output signal 
there is some uncertainty of what was originally sent. The conditional
entropy $H\left[{y\left(\omega\right)}|{x\left(\omega\right)} \right]$ 
at the output {\it for a given} 
$x\left(\omega\right)$
represents this uncertainty. 

The entropies  
$H\left[y\left(\omega\right)\right]$
and
$H\left[{y\left(\omega\right)}|{x\left(\omega\right)}\right]$ 
are defined in terms of the
corresponding distributions $p\left({y}\right)$ and
$p\left({y}|{x}\right)$ via the standard relation
\begin{eqnarray}
H & \equiv & - \int {\cal D}y\left(\omega\right)
p\left[y\left(\omega\right)\right]
\log\left(  p\left[y\left(\omega\right)\right] \right)
\label{eq:h}
\end{eqnarray}
where 
$ p \equiv p_y(y)$ 
for the entropy $H\left[y\left(\omega\right)\right]$, and
$ p \equiv p\left( y \left| \right. x \right)$ 
for the entropy 
$H\left[{y\left(\omega\right)}|{x\left(\omega\right)} \right]$, and
the functional integral is defined in the standard way
\begin{eqnarray}
\int {\cal D}\xi\left(\omega\right) \equiv
\lim_{M \to \infty}
c_M
\left[ \Pi_{m=1}^{M} \int d\xi\left(\omega_m\right) \right]
\end{eqnarray}
where is a normalization constant. 
 
For any communication link, the signal power 
is limited by the system hardware. Therefore, the maximum
of the mutual information in (\ref{eq:Shannon}) should be found under
the  constraint of the fixed total power $P_0$ at the input:
\begin{eqnarray}
P_0 & = &  \int {\cal D} x\left(\omega\right) 
\left| x\left(\omega\right) \right|^2 
p_x\left[  x\left(\omega\right)\right]
\end{eqnarray}

If the propagation in the communication
channel is described by a linear equation, then the input-output
relation for the system is given by
\begin{eqnarray}
y\left(\omega\right) = K\left(\omega\right) x\left(\omega\right)
+ n\left(\omega\right)
\end{eqnarray}
where $n\left(\omega\right)$ is the noise in the channel.
In this approximation, the problem of finding the maximum of the
mutual information can be solved exactly, with the corresponding 
input distribution $p_x$ being Gaussian\cite{Shannon}. 
If the amplifiers compensate exactly for the power losses in the fibers, 
the Channel Capacity is given by the Shannon formula (\ref{eq:Shannon}).

As follows from (\ref{eq:Shannon}), the better bit rates can
be obtained for the higher signal-to-noise ratio $P_0/P_N$. With this
in mind, the 
optics fiber communication systems are designed to operate
with the pulses of high power. As a result, the optics
fiber links operate in the regime, in which due to the Kerr
nonlinearity the refraction index of the
fiber strongly depends on the local electric field intensity. Therefore, 
a modern fiber optics communication system is, in fact, an essentially
nonlinear communication channel, and cannot be adequately described
within the framework of the Shannon's {\it linear} theory.

\begin{figure}
\begin{center} 
\leavevmode 
\epsfxsize 9 truecm
\epsfbox{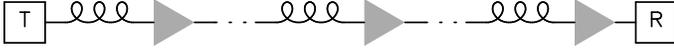} 
\end{center} 
\caption{
The schematical representation of a fiber optics communication channel
\protect\label{fig:schem}
}
\end{figure}

\section{The Model}

The first step in the calculation of the channel capacity is to find the
``input-output'' relation for the communication channel. The time evolution
of the electric field in the fiber $E(z,t)$, where $z$ is the distance
along the fiber, can be accuarately described in the 
``envelope approximation''\cite{Agrawal}, when 
\begin{eqnarray}
E\left(z,t\right) & = & A\left(z,t\right) \exp\left(i \left(\beta_0 z - 
\omega_0 t\right) \right) + {\rm c.c.}
\end{eqnarray}
where the function $A$ represents the slowly (compared to the light 
frequency) varying amplitude of the electric field
\vskip 3 truecm

\noindent
 in the fiber. The
evolution of $A\left(z,t\right)$ is 
described by the equation
\begin{eqnarray}
\frac{\partial A}{\partial z} +
\beta_1 \frac{\partial A}{\partial t} +
\frac{i}{2} \beta_2 \frac{\partial^2 A}{\partial t^2} +
\frac{\alpha}{2} A = 
i \gamma \left| A \right|^2 A
\label{eq:a}
\end{eqnarray}
Here the 
coefficients $\beta$ describe the frequency dependence of the wavenumber
\begin{eqnarray}
\beta\left({\omega}\right) & = & \beta_0 + \beta_1
\omega + 
\frac{\beta_2}{2} {\omega}^2  + O \left[{\omega}^3\right]
\end{eqnarray}
where $\omega$ is measured from the center of the band $\omega_0$.

The equation (\ref{eq:a}) neglects the effects such as the stimulated Raman
scattering and the stimulated Brillouin scattering\cite{Agrawal}, 
compared to the 
Kerr nonlinearity of the refraction index of the fiber, represented by the
term $\gamma \left| A \right|^2 A$.

The optical amplifiers incorporated into the communication system (see
Fig. 1) compensate for the power losses in the fiber, but 
due to spontaneous emission each of them will inevitably introduce 
noise $n(t) = \exp\left(i \omega_0 t \right) 
\int d\omega \ n_\omega \exp\left(i \omega t\right)$ into the channel.
Generally, even
in a single optical amplifier, the noise distribution at any
given frequency $\omega_0 + \omega$ within the channel bandwidth 
$n\left(\omega\right)$ is close to a Gaussian:
\begin{eqnarray}
p_n\left[n\left(\omega\right)\right] & \sim & \exp\left[
- \frac{\left|n\left(\omega\right)\right|^2}{P_N^\omega} \right]
\end{eqnarray}
This is even more so in a system with many independent amplifiers, 
due to the Central Limit Theorem.
For simplicity, the noise spectrum $P_N^\omega$ can be
assumed to be flat.

If the 
envelope function just before the amplifier is $A\left(t\right) \equiv
A^0_\omega \exp\left(i \omega t \right)$, then immediately after the
amplifier
\begin{eqnarray}
A_\omega & = & \exp\left(\frac{\alpha}{2} d\right) A^0_\omega  +
n_\omega
\label{eq:amplifier}
\end{eqnarray}
where $d$ is the span of a single fiber.

The equations (\ref{eq:a}), (\ref{eq:amplifier}) define the evolution
of the electric field envelope over one ``fiber-aplifier'' link of the
communication system. The total ``input-output'' relation will then
involve solving the corresponding equations for all $N$ iterations of the
single fiber-amplifier unit.

\section{The Perturbative Framework}

If one is able to calculate the ``output'' signal $y(\omega)$ in terms 
of the ``input'' $x(\omega)$ and the noise contributions of each of the 
amplifiers $n_{\omega}^{ \left\{\alpha\right\} }, \alpha = 1, \ldots, N$,
\begin{eqnarray}
y(\omega) = {\Phi} \left( {x(\omega)}; 
{n_\omega}^{ \left\{ 1 \right\} }, \ldots,
{n_\omega}^{ \left\{N \right\} } \right)
\label{eq:phi}
\end{eqnarray}
then the the
conditional distribution $p\left({y}|{x}\right)$ can be 
simply calculated as follows:
\begin{eqnarray}
p\left({y}|{x}\right) & = & 
\left\{ \Pi_{\alpha = 1}^{N-1}
\int {\cal D} {n_\omega}^{ \left\{\alpha\right\} } 
\ p_n\left[ {n_\omega}^{ \left\{\alpha\right\} } \right] \right\}
\ \nonumber \\
& \times & p_n\left[ {y(\omega)} - {\Phi}\left( {x(\omega)}; 
{n_\omega}^{ \left\{1\right\} }, 
\ldots, {n_\omega}^{ \left\{N\right\} } \right)
\right]
\label{eq:pxy_phi}
\end{eqnarray}
where $p_n$ is the distribution function of the noise, produced by 
a single amplifier. The output distribution $p_y\left({y}\right)$
 can then be directly related to the input distribution
$p_x\left({x}\right)$ via the standard relation
\begin{eqnarray}
p_y\left({y}\right) = \int {\cal D}{x(\omega)} \
p\left[{y(\omega)}|{x(\omega)}\right] \
p_x\left[{x(\omega)}\right] 
\label{eq:py}
\end{eqnarray}

Using Eqns. (\ref{eq:pxy_phi}),(\ref{eq:py}), one is able
to express the mutual information in terms of a single distribution
$p_x$. 
The calculation of the channel capacity then reduces to a standard
problem of finding the maximum of a (nontrivial) functional.

The equation (\ref{eq:a}) is, in fact, the well studied
\cite{NSE} nonlinear Shroedinger equation, with the time and distance
variables interchanged. Only some partial solutions of this equation are 
known, corresponding to solitons\cite{NSE,solitons}. However, in 
order to calculate the channel capacity, one
needs to find the general input-output relation for the communication system.
This implies solving a set of $N$ essentially nonlinear
equations (\ref{eq:a}) for {\it arbitrary} initial conditions.
Even knowing some partial solutions, doing such calculation exactly for
an essentially nonlinear system is not possible in a closed form.

In order to make progress, we note the presence of a natural perturbation
parameter in the problem, namely $\gamma$.In fact, the  fiber equation 
(\ref{eq:a}) is already an approximation, derived in the limit, 
when the change in the effective
refraction index due to pulse propagation, described by the 
nonlinear term $i \gamma \left| A \right|^2 A$, is {\it small}
compared to the ``unperturbed'' value of the index of refraction $n_0$.
We have developed a perturbative technique, when the 
solution of the nonlinear evolution equation, is represented as 
a power series in $\gamma$. Solving (\ref{eq:a}) separately
for each power of $\gamma$, and using (\ref{eq:amplifier}), for the
input-output relation of a single fiber-amplifier unit
${ \Phi}_\omega^{(n)}$, defined as
\begin{eqnarray}
A_\omega^{(n)} & = & { \Phi}_\omega^{(n)}\left(A_\omega^{(n-1)}\right),
\end{eqnarray}
we obtain:
\begin{eqnarray}
{ \Phi}_\omega^{(n)}\left({A_\omega}^{(n-1)}\right) & = &
\left[ A_\omega^{\left(n-1\right)} + \sum_{\ell = 1}^{\infty}
\gamma^\ell 
{\cal F}_\omega^{(\ell)}\left({A_\omega^{\left(n-1\right)} }\right) \right]
\nonumber \\
& \times & 
\exp\left(
 - i \kappa_\omega d \right) + n_\omega
\label{eq:phi_pert}
\end{eqnarray}
where $d$ is the length of a single fiber,
\begin{eqnarray}
\kappa_\omega & = & 
 \beta_1 \omega - \frac{1}{2} \beta_2 \omega^2 
\label{eq:kappa} 
\end{eqnarray}
The procedure for the calculation of the functions  
${\cal F}_\omega^{(\ell)}$, described in detail 
Appendix \ref{sec:App1}, can be carried to an
arbitrary order $\ell$.

The further calculation then involves the following 
steps:
\begin{itemize}
\item{ Iterating Eq. (\ref{eq:phi_pert}) $N$ times, to
 obtain the ``input-output'' relation for the whole communication 
system $\Phi_\omega\left[x\left(\omega\right); n_\omega^{(1)}, \ldots,
n_\omega^{(N)} \right]$}
\item{substituting the result into Eqns. (\ref{eq:pxy_phi}),
(\ref{eq:py}) to obtain the conditional distribution $p(x|y)$ 
and the output distribution $p_y(y)$ in terms of the input distribution
$p_x(x)$ as expansions in powers of $\gamma$}
\item{calculating the entropies $H\left[y\left(\omega\right)\right]$
and
$H\left[{x\left(\omega\right)}|{y\left(\omega\right)}\right]$, and
the mutual information $R$}
\end{itemize}

Following these steps, the calculation of the channel capacity becomes
a straightforward procedure. In Appendix \ref{sec:App2} we describe it in
detail, using a simple nonlinear channel $y\left(\omega\right) 
= x\left(\omega\right) \exp\left( - \phi\left[ x\left(\omega\right)\right]
\right) + n\left(\omega\right)$ as an example.

\section{The Fiber Link Channel Capacity}

After a tedious, but straigthforward calculation, we obtain:
\begin{eqnarray}
H\left[y\left(\omega\right)\right] & = & H_0\left[y\left(\omega\right)\right]
- \Delta C_1 - \Delta H_y + {\cal O}\left(\gamma^4\right)
\label{eq:h_y}
\end{eqnarray}
and
\begin{eqnarray}
H\left[{y\left(\omega\right)}|{x\left(\omega\right)}\right]
 & = & H_0\left[{y\left(\omega\right)}|{x\left(\omega\right)}\right]
+ \Delta C_2 + {\cal O}\left(\gamma^4\right)
\label{eq:h_xy}
\end{eqnarray}
where $H_0\left[y\right]$ and  $H_0\left[x|y\right]$ are given
by the standard expressions for a linear channel\cite{Shannon}.
In the limit of large signal-to-noise ratio $P_0 \gg P_N$ we obtain:
\begin{eqnarray}
\Delta{C}_1 & = &  N^2 W  \left(\frac{\gamma P_0}{\alpha}\right)^2 
Q_1\left(\alpha d, \frac{\beta_2^2 W^4}{\alpha^2} \right)
\label{eq:deltaC1}
\\
\Delta{C}_2 & = &  \frac{4}{3} 
\left(N^2 - 1\right) W  \left(\frac{\gamma P_0}{\alpha}\right)^2 
Q_2\left(\alpha d, \frac{\beta_2^2 W^4}{\alpha^2} \right)
\label{eq:deltaC2}
\end{eqnarray}
and the functions $Q_1$ and $Q_2$ are defined as follows:
\begin{eqnarray}
Q_1\left(u,z\right) & = & \int_{-1/2}^{1/2} dx_1
\int_{-1/2}^{1/2} dx_2 \int_{1/2}^{\bar{x}} dx
\ \nonumber \\
& \times & f\left(u, z; x_1, x_2, x\right)
\\
Q_2\left(u,z\right) & = & \int_{-1/2}^{1/2} dx_1
\int_{-1/2}^{1/2} dx_2 \int_{-1/2}^{1/2} dx
\ \nonumber \\
& \times & f\left(u, z; x_1, x_2, x\right)
\end{eqnarray}
where $\bar{x} \equiv {\rm max}\left[1/2,1/2+x_1+x_2\right]$, and
\begin{eqnarray}
f\left(u, z; x_1, x_2, x\right) \equiv 
\frac{\left|1 - 
\exp\left( - u - i v^2
\right) \right|^2  
}{1+v^2}
\end{eqnarray}
where
\begin{eqnarray}
v \equiv z \left(x-x_1\right) \left(x-x_2\right)
\end{eqnarray}
The correction 
\begin{eqnarray}
\Delta H_y & = & \gamma^2 
W \int {\cal D}{y\left(\omega\right)} 
p_y^{(0)}\left[y\left(\omega\right)\right] 
\left( p_y^{(1)}\left[y\left(\omega\right)\right] 
\right)^2
\label{eq:hy_1}
\end{eqnarray}
is caused by the deviations of the otput distribution 
\begin{eqnarray}
p_y & = & p_y^{(0)} + \sum_\ell \gamma^\ell p_y^{(\ell)}
\end{eqnarray}
from the Gaussian form 
\begin{eqnarray}
p_y^{(0)}\left[y\left(\omega\right)\right] \sim 
\exp\left( - 
\frac{\left|y\left(\omega\right)\right|^2}{P_\omega + P^N_\omega} \right)
\end{eqnarray}
where $P_\omega$ (such that $P_0 = \int d\omega P_\omega$) 
is the input power at frequency $\omega$:
\begin{eqnarray}
p_x^{(0)}\left[x\left(\omega\right)\right] \sim 
\exp\left( - 
\frac{\left|x\left(\omega\right)\right|^2}{P^0_\omega} \right)
\end{eqnarray}
Note, that the correction $\Delta H_y \geq 0$ and equals to zero
only when the distribution $p_y^{(1)} = 0$.
Therefore, as follows from Eqns. (\ref{eq:h_y}),(\ref{eq:hy_1}), 
in the second order in nonlinearity $\gamma$ the mutual information
$R$ has the maximum, when $p_y^{(1)} = 0$, or, equivalently,
when the {\it output} distribution is Gaussian up to the {\it first}
order in nonlinearity.

For a general nonlinear channel, that would correspond to the input
distribution, being non-Gassian already in the first order in 
$\gamma$. The corresponding correction can be obtained from
Eq. (\ref{eq:py}), taken only up to the first orded in nonlinearity:
\begin{eqnarray}
\left.
\frac{\partial}{\partial \gamma}\left[
\int {\cal D}{x(\omega)} \
p\left[{y(\omega)}|{x(\omega)}\right] \
p_x\left[{x(\omega)}\right] \right] \right|_{\gamma = 0}
& = & 0
\label{eq:py1}
\end{eqnarray}
Generally, such an integral would yield $p_x\left(x\right)
= p_x^{(0)}\left(x\right) 
\left(1 + \gamma p_x^{(0)}\left(x\right) \right)$,
where $p_x^{(0)}$ is Gaussian, and $p_x^{(1)} \neq 0$.
However, it is straightforward to show, that 
for the fiber optics channel described by Eq.
(\ref{eq:a}), a Gaussian input distribution leads to 
non-Gaussian corrections in the output distribution
starting only from the {\it second} order. Therefore,
the requrement $p_y^{(1)} =0$ is satisfied, when
the input distribution is such that $p_x^{(1)} = 0$.

For the channel capacity, defined as the maximum value of the
mutual information, we obtain:
\begin{eqnarray}
C & = & W \log\left(1  + \frac{P_0}{P_N}\right) - 
\Delta{C}_1 - \Delta{C}_2 
+ {\cal O}\left(\gamma^4\right)
\label{eq:cpert}
\end{eqnarray} 
The equation (\ref{eq:cpert}) yields the result for the
 fiber optics channel capacity in the second order in the nonlinearity
$\gamma$. In the next section we will discuss the physical origins
of the corrections $\Delta C_1$ and $\Delta C_2$.

\begin{figure}
\begin{center} 
\leavevmode 
\epsfysize = 5.cm 
\epsfbox{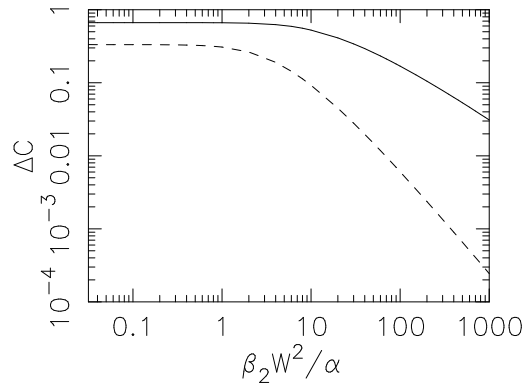} 
\end{center} 
\caption{The corrections to the channel capacity, $\Delta C_1$ and
$\Delta C_2$, in units of $W N^2 \gamma^2 P_0^2 \alpha^{-2}$, shown
as functions of  $|\beta_2| W^2/\alpha$
in the limit $\alpha d \gg 1$, $N \gg 1$. The correction
$\Delta C_2$ is represented by the solid line, while
$\Delta C_1$ corresponds to the dashed line. Note, that $\Delta C_1$, which 
describes the effect of the 
power leakage from  the bandwidth, is more strongly affected by the
dispersion
\label{fig:dC}
}
\end{figure}

\section{The  Discussion and the Conclusions}

In the spirit  of the Shannon formula, the decrease of the 
capacity of a communication channel with a fixed bandwidth can
be attributed to (i) the effective suppression of the signal
power, and (ii) the enhancement of the noise. The corrections to the
channel capacity, derived in the previous section, can be interpreted
as resulting precisely from these two effects.

The four-wave   
scattering\cite{Agrawal}, induced  by the
fiber nonlinearity, inevitably leads to the processes, which generate
photons with the frequencies  outside the channel bandwidth.
Such photons, are not recorded by the ``receiver'', and are lost for
the purpose of the information transmission.
 This 
corresponds  to
an effective  bandwidth power dissipation, and should therefore lead
to a decrease of the channel capacity. Since for small nonlinearity this
power loss $\Delta P \sim \gamma^2$, the dimension analysis implies
\begin{eqnarray}
\Delta P & = &  \frac{ \gamma^2 P_0^2}{\alpha^2} 
F\left(\frac{\beta W^2}{\alpha}\right) 
\
\end{eqnarray}
Such scattering processes are suppressed, when the scattering leads to a 
subsantial change of the total momentum $\delta \kappa\left[ \omega_1,
\omega_2 \rightarrow \omega_3, \omega\right]$, so that the corresponding
scattering rate 
\begin{eqnarray}
S\left[ \omega_1,
\omega_2 \rightarrow \omega_3, \omega\right]
& \sim & 
\frac{\delta\left(\omega_1+\omega_2 - \omega_3- \omega\right)}
{1 + \left(\delta\kappa/\kappa_0\right)^2}
\label{eq:S}
\end{eqnarray}
In the spirit of the uncertainty realtion, 
$\kappa_0 \sim 1/L_{\rm eff}$, where 
$L_{\rm eff} $ corresponds to the length of the concentration 
of the power of the signal in the fiber. For a small absorption
coefficient $\alpha \ll 1/d$ the distance $L_{\rm eff} $ is of
the order of the fiber length $d$, while in the opposite limit
$\alpha \gg 1$ the effective length $L_{\rm eff} \approx 1/\alpha $.

Using Eqn. (\ref{eq:kappa}), and the energy conservation
$\omega_3 = \omega_1 + \omega_2 - \omega$, the momentum change
$\delta \kappa\left[ \omega_1,
\omega_2 \rightarrow \omega_3, \omega\right]$
can be expressed as
\begin{eqnarray}
\delta \kappa & = & \beta_2 \left(\omega - \omega_1\right)
\left(\omega - \omega_1\right)
\label{eq:dk}
\end{eqnarray}
Substituting (\ref{eq:dk}) into (\ref{eq:S}), for the 
channel capacity loss due to the bandwidth power ``leakage'',
in the limit $P_0 \gg P_N$, and $\alpha d \gg 1$, we obtain
\begin{eqnarray}
\Delta C_P & \sim & W \frac{\Delta P}{P} \sim 
W \frac{\gamma^2 P_0^2}{\alpha^2} {\int}_W d\omega_1 \ 
\int_W d\omega_2
 \ \int_W d\omega_3 \nonumber \\
& \times &
\int_{\omega \notin W} d\omega \ 
S\left[ \omega_1, \omega_2 \rightarrow \omega_3, \omega\right]
\nonumber \\
& = & W \frac{\gamma^2 P_0^2}{\alpha^2}
 \int_W d\omega_1 \ \int_W d\omega_2 
\int_{\omega \notin W} d\omega \nonumber \\
& \times & 
\frac{1}{1 + \left(\beta_2/\alpha\right)^2
\left(\omega - \omega_1\right)^2 \left(\omega - \omega_2\right)^2}
\label{eq:CPpert}
\end{eqnarray}
which in the appropriate limit is consistent with $\Delta C_1$.

In Fig. 2 we plot the dependence of $\Delta C_1$ on the dimensionless
parameter $\beta_2 W^2 /\alpha$. Since momentum change $\delta \kappa$
is proportional to $\beta_2$, the increase of the dispersion leads to
a strong suppression of the power leakage from the bandwidth window,
and of the corresponding correction to the channel capacity.

In a communication system with many ``fiber-amplifier'' units, the
fiber nonlinearity leads not only to the mixing of the signals at
different frequencies, but also to the mixing of the signal with 
the noise. Qualitatively, this would correspond to an effective
enhancement of the noise power in the system, and therefore to a loss
of the channel capacity. This effect is not present, when the system
has only one ``fiber-amplifier'' link, which explains the 
appearance of the $(N-1)$ factor in $\Delta C_2$ and $\Delta C_3$.

The effective noise enhancement is caused by the scattering processes,
which involve a ``signal photon'' and a photon, produced due to 
spontaneous emission in one of the amplifiers. The total power 
of this extra noise can be expressed as
\begin{eqnarray}
\frac{\Delta P_N}{P_0} & \sim & \gamma^2 
\frac{P_0 P_N}{\alpha^2} 
\int_W d\omega_1 \ \int_W d\omega_2
\ \int_W d\omega_3
\int_{W} d\omega \ \nonumber \\ 
& \times & S\left[ \omega_1, \omega_2 \rightarrow \omega_3, \omega\right]
\end{eqnarray}
The corresponding correction to the capacity
\begin{eqnarray}
\Delta C_N & \sim & W \frac{\Delta P_N}{P_N} \sim 
W \frac{\gamma^2 P_0^2}{\alpha^2}
 \int_W d\omega_1 \ \int_W d\omega_2
\int_{ W} d\omega \ \nonumber \\
& \times & 
\frac{1}{1 + \left(\beta_2/\alpha\right)^2
\left(\omega - \omega_1\right)^2 \left(\omega - \omega_2\right)^2}
\label{eq:CNpert}
\end{eqnarray}
where we assumed $\alpha d \gg 1$. In this limit (\ref{eq:CNpert})
is up to a constant factor identical to $\Delta C_2$. 

The dependence
of $\Delta C_2$ on $\beta_2 W^2/\alpha$ is also shown in Fig. 2. Note,
that $\Delta C_2$ also decreases with the increase of the 
dispersion, but more slowly than  $\Delta C_1$. Since the scattering
processes, which contribute to $\Delta C_1$, need to ``move'' one of
the frequencies out of the bandwidth window, they generally involve
a substantial change of the total momentum, and are therefore more 
strongly affected by the dispersion.

The two physical effects, described above, determine the 
fundamental limit to the bit rate for a fiber optics communication
system. As follows from our analysis (see Fig. 2),
the relative contributions of $\Delta C_1$ and $\Delta C_2$,
often referred to as the ``four-wave mixing'', can be suppressed
by choosing a fiber with a large dispersion, or when using a larger
bandwidth. 

In our analysis, we treated the whole available bandwidth as a single
channel. As a result, the cross-phase modulation\cite{Agrawal}, which
severely limits the performance of advanced wavelength-division
multiplexing systems\cite{Agrawal2} (WDM), does not affect the 
channel capacity. The reason for this seemingly contradictory
behaviour, is that in a WDM system, the ``receiver'', tuned to a 
particular WDM channel, has no information on the signals at the 
other channels. Therefore, even in the absense of the ``geniune''
noise, the nonlinear interaction between different channels,
leading to a change in the signal in any given channel, will be 
 an effective noise source, thus limiting the communication rate.
This limit however is not fundamental, and can be overcome by using
the whole bandwidth all together.

In conclusion, we developed a perturbative method for the 
calculation of the channel capacity for fiber optics communication 
systems. We obtained analytical expressions for the corrections
to the Shannon formula due to fiber nonlinearity. We have shown that,
 compared to the 
Shannon limit, the actual channel 
capacity is substantially suppressed by the photon scattering processes,
caused by the fiber nonlinearity.

\appendix

\section{Perturbative Solution of the Propagation Equation}
\label{sec:App1}

In this Appendix we describe the perturbative solutuion of the
nonlinear equation (\ref{eq:a}) with the boundary condition
\begin{eqnarray}
A\left(0,t\right) & = & \int d\omega \ x\left(\omega\right) 
\exp\left(i \omega t\right)
\label{eq:initial}
\end{eqnarray}
We represent $A\left(z,t\right)$ as a power series
\begin{eqnarray}
A\left(z,t\right) & = & \int d\omega 
\exp\left(i \omega t - 
\left[\frac{\alpha}{2} + i \kappa_\omega\right] z\right)
\sum_{n = 0}^\infty \gamma^\ell \nonumber \\
& \times & {\cal F}_\ell\left(z,\omega\right)
\label{eq:anzats}
\end{eqnarray}
where $\kappa_\omega$ is defined is (\ref{eq:kappa}). Substituting
(\ref{eq:anzats}) into Eqns. (\ref{eq:a}),(\ref{eq:initial}), we 
obtain:

\noindent
(i) for $\ell = 0$
\begin{eqnarray}
\frac{\partial {\cal F}_0\left(z,\omega\right)}{\partial z} & = & 0
\\
{\cal F}_0\left(0,\omega\right) & = & x\left(\omega\right)
\end{eqnarray}

\noindent
(ii) for $\ell \neq 0$
\begin{eqnarray}
\frac{\partial {\cal F}_\ell\left(z,\omega\right)}{\partial z} & = & 
i \sum_{\ell_1, \ell_2, \ell_3 = 1}^{\ell-1}
 \delta_{\ell - 1, \ell_1 + \ell_2 + \ell_3} 
\int d\omega_1 \int d\omega_2 \ \nonumber \\
& \times &
{\cal F}_{\ell_1}\left(z,\omega_1\right)
{\cal F}_{\ell_2}\left(z,\omega_2\right)
{\cal F}^*_{\ell_3}\left(z,\omega_1+ \omega_1- \omega\right)
\nonumber \\
& \times & 
\exp\left[ i \left(\kappa_{\omega_1} + \kappa_{\omega_1}
- \kappa_{\omega_1+ \omega_2 - \omega} - \kappa_{\omega} \right) \right]
\\
{\cal F}_\ell\left(0,\omega\right) & = & 0
\end{eqnarray}
where $\delta$ in the Kronekker's delta-function.

For any $\ell$, these equations reduce to {\it linear} first order differential
equation, and can be solved straightforwardly. For example, the solutions 
for the first three terms in the anzats (\ref{eq:anzats})  are given by:
\begin{eqnarray}
{\cal F}_0\left(z,\omega\right) & = & x\left(\omega\right) \\
{\cal F}_{1}\left(z,\omega\right) & = &
\int d\omega_1 \int d\omega_2 \ 
F^\omega_{\omega_1 \omega_2} 
x\left(\omega_1\right) x\left(\omega_2\right) 
\nonumber \\
& \times & x^*\left(\omega_1 + \omega_2 - 
\omega\right)
 \\
{\cal F}_{2}\left(z,\omega\right) & = &
\int d\omega_1 \int d\omega_2 \int d\omega_3 \int d\bar{\omega}
 \left[ G^\omega_{\omega_1 \omega_2 \omega_3 \bar{m}} \right.
\nonumber \\
& \times & x^*\left(\omega_1\right)
x^*\left(\omega_2\right) x\left(\omega_1 + \omega_2 - \bar{\omega}\right)
x\left(\omega_3\right) 
\nonumber \\
& \times & x\left(\omega + \bar{\omega} - \omega_3\right) + 
H^\omega_{\omega_1 \omega_2 \omega_3 \bar{\omega}} 
x\left(\omega_1\right) x\left(\omega_2\right) x\left(\omega_3\right)
\nonumber \\
& \times & 
\left. 
x^*\left(\omega_1 + \omega_2 - \bar{\omega}\right) 
x^*\left(\bar{\omega} + \omega_3 - \omega\right) \right]
\end{eqnarray}
Here the functions $F$, $G$ and $H$ are given by
\begin{eqnarray}
F^\omega_{\omega_1 \omega_2} & = & 
 i 
\frac{1 - \exp\left( - \alpha d - i 
\phi^\omega_{\omega_1 \omega_2} d \right)}
{ \alpha + i \phi^\omega_{\omega_1 \omega_2} }
\nonumber \\
G^m_{\omega_1 \omega_2 \omega_3 \bar{\omega}} & = & - 
\left(F^{\bar{\omega}}_{\omega_1 \omega_2}\right)^*
\left(F^{{\omega_3}}_{\omega \bar{\omega} }\right)^* 
\nonumber \\
& \times & 
\left(1 - 
\exp\left( - \alpha d + 
           i \phi^{\omega_3}_{\omega \bar{\omega} }d \right)
\right)
\nonumber \\
& - & \left( \frac{  F_{\omega_1 \omega_2}^{\bar{\omega}}  }{
\frac{1}{ F_{\omega_1 \omega_2}^{\bar{\omega}} } + 
\frac{1}{ F_{\omega \bar{\omega}}^{\omega_3} } }
\right)^*
\nonumber \\
& \times &  \left( 1 - \exp\left( - 2 \alpha d 
+ \left(
\phi^{\bar{\omega}}_{\omega_1 \omega_2} + 
\phi^{\omega_3}_{\omega \bar{\omega}} 
 \right) d \right)
\right)
\nonumber \\
H^\omega_{\omega_1 \omega_2 \omega_3 \bar{\omega}} & = & 
2
\left(F^{\bar{\omega}}_{\omega_1 \omega_2}\right)
\left(F^{{m}}_{\bar{\omega} \omega_3}\right)
\nonumber \\
& \times & 
\left(1 - 
\exp\left( - \alpha d - 
           i \phi^{\omega}_{\bar{\omega} \omega_3} d \right)
\right)
\nonumber \\
& -  & 2 \left( \frac{  F_{\omega_1 \omega_2}^{\bar{\omega}}  }{
\frac{1}{ F_{\omega_1 \omega_2}^{\bar{\omega}} } + 
\frac{1}{ F_{\bar{\omega} \omega_3}^{\omega} } }
\right)^* 
\nonumber \\
& \times & \left( 1 - \exp\left( - 2 \alpha d -  i 
\left( 
\phi^{\bar{\omega}}_{\bar{\omega_1} \omega_2}
+ 
\phi^{\omega}_{\bar{\omega} \omega_3}
\right)
d
\right)
\right)
\nonumber 
\end{eqnarray}
where
\begin{eqnarray}
\phi^\omega_{\omega_1 \omega_2} & = & 
\left(\kappa_{\omega_1} + 
\kappa_{\omega_2}
- \kappa_{\omega_1 + \omega_2 - \omega} - \kappa_\omega \right)
\nonumber
\end{eqnarray}

\section{The Perturbative Calculation of the Channel Capacity}
\label{sec:App2}

In this Appendix, we describe the perturbative calculation of the
capacity of the simple nonlinear channel
\begin{eqnarray}
y\left(\omega\right) 
= x\left(\omega\right) \exp\left( i \gamma 
\phi\left[ x\left(\omega\right)\right]
\right) + n\left(\omega\right)
\label{eq:app_y}
\end{eqnarray}
Here $\phi\left[ x\right]$ is an arbitrary real function, and
the noise $n\left(\omega\right)$ is a Gaussian random variable:
\begin{eqnarray}
p_n\left[n\left(\omega\right)\right] & \sim & \exp\left[
- \frac{\left|n\left(\omega\right)\right|^2}{P_N} \right]
\label{eq:app_pn}
\end{eqnarray}
The fact, that the noise in this model is additive, implies that the
conditional distribution $p\left(y|x\right)$ is fixed, and defined
by the noise distribution $p_n$:
\begin{eqnarray}
p\left(y\left|\right.x\right) & = &p_n\left[
y - x \exp\left( i \gamma \phi\left( x\right)\right) \right]
\label{eq:app_pyx}
\end{eqnarray}
It is therefore straightforward to show, that the entropy
$H\left[{y}|{x}\right]$ does not depend on $\gamma$:
\begin{eqnarray}
H\left[{y}|{x}\right] & = & - 
\int dy p_n\left(y - x e^{i \gamma \phi(x)}\right)
\nonumber \\
& & \times
\log\left[p_n\left(y - x e^{i \gamma\phi(x)}\right)\right]
\nonumber \\
& = &  - \int dz p_n\left(z\right)
\log\left[p_n\left(z\right)\right] \nonumber 
\\
& = & \log\left[2 \pi e P_N\right]
\label{eq:app_hyx}
\end{eqnarray}
In order to calculate the entropy $H\left[{y}\right]$, we represent the
output distribution as a power series in $\gamma$:
\begin{eqnarray}
p_y\left(y\right) & = & p_y^0\left(y\right) \left[1 + 
\sum_{n=1}^\infty p_y^{(n)}\left(y\right) \right]
\label{eq:app_py}
\end{eqnarray}
where $p_y^0(y)$ is the ``unperturbed'', Gaussian distribution
\begin{eqnarray}
p_y^0\left(y\right) & = & \frac{1}{\pi \left(P_0 + P_N\right)}
\exp\left( - \frac{\left| y \right|^2}{P_0 + P_N} \right)
\label{eq:app_p0y}
\end{eqnarray}
corresponding to the linear channel $y = x + n$. Substituting 
(\ref{eq:app_py}) into the definition of the entropy $H_y$, 
Eq. (\ref{eq:h}), we obtain:
\begin{eqnarray}
H_y & = & \log\left[2 \pi e \left(P_0 + P_N\right)\right]
- \frac{1}{2} \int p_y^{(1)}\left(y\right)^2 p_y^0\left(y\right) 
\nonumber \\
& + & \frac{1}{P_0 + P_N} \left[\int dy \left|y\right|^2 p_y\left(y\right)
\right. \nonumber \\
& - & \left. \int dy \left|y\right|^2 p_y^0\left(y\right)\right]
\label{eq:app_hy}
\end{eqnarray}
The second term in Eq. (\ref{eq:app_hy}), 
$(1/2)\int p_y^{(1)}\left(y\right)^2 p_y^0\left(y\right)  $, 
represents the difference of the output distribution
from Gaussian, and corresponds to the contribution
$\Delta H_y$ in Eq. (\ref{eq:h_y}). Note, that in the second order in
nonlinearity the deviations of the output distribution from Gaussian lead
to a {\it decrease} of capacity.

The third term, 
$(1/(P_0 + P_N))\left[\int dy \left|y\right|^2 p_y\left(y\right)
- \int dy \left|y\right|^2 p_y^0\left(y\right)\right] $, 
is proportional to the change of the 
output power, $\int dy \left| y\right|^2 p_y\left(y\right)$, due
to nonlinearity,
and corresponds to $\Delta C_1$ in Eq. (\ref{eq:h_y}). Generally,
the nonlinearity leads to energy exchange between different degrees of
freedom in the channel (e.g. between different frequencies), and to the
power leakage out of the bandwidth window. However, for the specific
(nad non-generic) example, chosen in the present Appendix, this
exchange is absent, since the output power
\begin{eqnarray}
\langle \left| y \right|^2 \rangle & = & 
\langle \left| x \right|^2 \rangle +
\langle \left| n \right|^2 \rangle = P_0 + P_N
\label{eq:app_power}
\end{eqnarray}
does not depend on the nonlinearity.

Substituting (\ref{eq:app_power}) in Eq. (\ref{eq:app_hy}), and using 
Eq. (\ref{eq:app_hyx}), for the mutual information $R$ we obtain:
\begin{eqnarray}
R & = & \log\left[1 + \frac{P_0}{P_N}\right]
- \frac{1}{2} \int p_y^{(1)}\left(y\right)^2 p_y^0\left(y\right) 
\label{eq:app_R}
\end{eqnarray}
As immediately follows from Eq. (\ref{eq:app_R}), the channel capacity,
equal to the maximum of the mutual information, is given by the
Shannon formula (\ref{eq:Shannon}), and is achieved when 
\begin{eqnarray}
p_y^{(1)}\left(y\right) & = & 0
\label{eq:app_py_max}
\end{eqnarray}

The next step is to calculate the {\it input} distribution 
\begin{eqnarray}
p_x\left(x\right) & = & p_x^0\left(x\right) \left[1 + 
\sum_{n=1}^\infty p_x^{(n)}\left(x\right) \right]
\label{eq:app_px}
\end{eqnarray}
corresponding to (\ref{eq:app_py_max}). The general relation between
the input and the output distributions is defined by the conditional
distribution $p(y|x)$:
\begin{eqnarray}
p_y\left({y}\right) = \int d{x} \
p\left({y}|{x}\right) \
p_x\left({x}\right) 
\label{eq:app_py_px}
\end{eqnarray}
and, considered as an equation for $p(x)$, is a Fredholm intergal
equation of the first kind. Note however, that since Eq. (\ref{eq:app_py_max})
represents not the whole output distribution, but only it's first order
term $p^{(1)}_y(y)$, we can expand Eq. (\ref{eq:app_py_px}) and keep only
the terms up to the first order in $\gamma$. We obtain:
\begin{eqnarray}
&  & \int d{x} \
\left.p\left({y}|{x}\right)\right|_{\gamma=0} \
 \ p_x^0\left({x}\right) \ p_x^{(1)}\left({x}\right)
\nonumber \\
& + & 
 \int dx \
\left. 
\frac{\partial}{\partial \gamma}p\left(y|x\right)\right|_{\gamma=0}
 \ p_x^0\left({x}\right) 
 = 0,
\label{eq:app_py_px_1}
\end{eqnarray}
Substituting here the conditional distribution
from Eq. (\ref{eq:app_pyx}), we obtain:
\begin{eqnarray}
\int dx \ p_n\left(y - x\right) p_x^0\left(x\right) p_1\left(x\right)
= \frac{i}{P_N} \nonumber \\
\times  \int dx \ p_n\left(y - x\right) p_x^0\left(x\right)
\phi\left(x\right) \left(x^* y - y^* x\right) ,
\label{eq:app_py_px_2}
\end{eqnarray}
Using the identity
\begin{eqnarray}
y p_n\left(y - x\right) & = & 
\left(x + P_N \frac{\partial}{\partial x^*}\right) p_n\left(y - x\right),
\end{eqnarray} 
and integrating by parts, we can represent the right hand side of
(\ref{eq:app_py_px_2}) as follows:
\begin{eqnarray}
i \int dx \ p_n\left(y - x\right) p_x^0\left(x\right)
\frac{i}{P_N}\phi\left(x\right) \left(x^* y - y^* x\right)  = 
\nonumber \\ 
i \int dx \  p_n\left(y - x\right) p_x^0\left(x\right)
\left( 
x^* \frac{\partial \phi\left(x\right)}{\partial x^*} 
-
x \frac{\partial \phi\left(x\right)}{\partial x} 
\right)
\label{eq:app_rhs}
\end{eqnarray}
Therefore, as follows from Eqns. (\ref{eq:app_py_px_2}) and 
(\ref{eq:app_rhs}), the input distribution
\begin{eqnarray}
p_x^{(1)}\left(x\right) & = & 
i
\left( 
x^* \frac{\partial \phi\left(x\right)}{\partial x^*} 
-
x \frac{\partial \phi\left(x\right)}{\partial x} 
\right)  
\label{eq:app_px_max}
\end{eqnarray}

This procedure can be followed up for all orders in $\gamma$. By a 
direct calculation, it is straightforward to show, that the 
channel capacity is represented by the Shannon result (\ref{eq:Shannon}),
which is achieved when for any $n > 1$
\begin{eqnarray}
\left\{
\begin{array}{ll}
p_x^{(n)}\left(x\right) =  0 \\
p_y^{(n)}\left(y\right) =  0
\end{array}
\right.
\label{eq:app_higher_orders}
\end{eqnarray}
Subsitituting (\ref{eq:app_higher_orders}) and 
(\ref{eq:app_px_max}) into Eq. (\ref{eq:app_px}), for the input distibution
we finally obtain:
\begin{eqnarray}
p_x\left(x\right) & = & \frac{1}{\pi P_0}
\left[1 + i \gamma 
\left( 
x^* \frac{\partial \phi\left(x\right)}{\partial x^*} 
-
x \frac{\partial \phi\left(x\right)}{\partial x} 
\right)
\right]
\nonumber \\
& \times & 
\exp\left[
- \frac{\left|x\right|^2}{P_0}
\right]
\label{eq:app_px_total}
\end{eqnarray}
with the corresponding channel capacity
\begin{eqnarray}
C & = & W \log\left[ 1 + \frac{P_0}{P_N} \right]
\label{eq:app_Shannon}
\end{eqnarray}

This result has a simple physical meaning. When the input distribution
is organized in such a way, that the quantity 
$z = x \exp\left( i \gamma \phi\left(x\right)\right)$ has the Gaussian 
distribution, then, considering $z$ as input, the communication
channel becomes linear: $y = z + n$, and the channel capacity is therefore
given by the Shannon formula (\ref{eq:Shannon}),(\ref{eq:app_Shannon}). 
The corresponding input distribution is then defined by the
Jacobian of the transformation from $x \equiv x_R + i x_I$ to 
$z \equiv z_R + i z_I$, 
$\partial\left(z_R, z_I\right)/\partial\left(x_R, x_I\right)$ (note, that 
$x_R$, $x_I$, $z_R$, $z_I$ are defined as {\it real} variables):
\begin{eqnarray}
p_x\left(x\right) & = & 
\frac{1}{\pi P_0}
\frac{\partial\left(z_R, z_I\right)}{\partial\left(x_R, x_I\right)}
\exp\left[ - \frac{\left|x_R^2 + x_I^2\right|^2}{P_0} \right]
\label{eq:app_pzpx}
\end{eqnarray}
which reduces to the distribution (\ref{eq:app_px_total}), since
\begin{eqnarray}
 \frac{\partial\left(z_R, z_I\right)}{\partial\left(x_R, x_I\right)} 
& = & 1 + \gamma x_R \frac{\partial \phi\left(x_R, x_I\right)}{\partial x_I}
-  \gamma x_I \frac{\partial \phi\left(x_R, x_I\right)}{\partial x_R}
\nonumber \\
& \equiv & 1 + i \gamma 
x^* \frac{\partial \phi\left(x,x^*\right)}{\partial x^*} 
- i \gamma
x \frac{\partial \phi\left(x,x^*\right)}{\partial x} 
\end{eqnarray}

This result should be contrasted to the so called ``Gaussian estimate'' of 
the channel capacity\cite{Telatar99}. In the latter appoach, the
information channel is described by the {\it joint Gaussian} distribution
\begin{eqnarray}
{\cal P}\left(x_\omega,y_\omega\right) \sim
\exp\left( - \left[ x_\omega^* \ y_\omega^*\right] {\cal A} 
\left[
\begin{array}{c}
x_\omega \\
y_\omega
\end{array}
\right]
\right)
\label{eq:app_joint}
\end{eqnarray}
where
\begin{eqnarray}
{\cal A} & = & 
\left[
\begin{array}{cc}
\langle x_\omega^* x_\omega\rangle & \langle x_\omega^* y_\omega\rangle \\
\langle y_\omega^* x_\omega\rangle & \langle y_\omega^* y_\omega\rangle
\end{array}
\right]^{-1}
\end{eqnarray}
The channel capacity is then estimated as the mutual information, 
corresponding to the distribution (\ref{eq:app_joint}):
\begin{eqnarray}
C_G = \int d\omega \log\left[
\frac{ \langle x_\omega^* x_\omega\rangle 
\langle y_\omega^* y_\omega\rangle}
{\langle x_\omega^* x_\omega\rangle \langle y_\omega^* y_\omega\rangle
- \langle x_\omega^* y_\omega\rangle \langle y_\omega^* x_\omega\rangle}
\right]
\label{eq:app_Cg}
\end{eqnarray}
Under the constraint of the fixed input power $\int d\omega \langle 
\left| x_\omega \right|^2 \rangle$, the estimate 
(\ref{eq:app_Cg}) was shown\cite{Telatar99} to 
give the low bound to the channel capacity.

For the model channel considered in the present Appendix, the ``Gaussian
estimate'' yields an expression, 
{\it different} from the Shannon result. For example,
when $\phi\left(x\right) = \left| x \right|^2$, we obtain
\begin{eqnarray}
C_G & = & - W \log\left[1 - 
\frac{P_0}{\left(P_0 + P_N\right)\left(1 + \gamma^2 P_0^2\right)^2}
\right] \nonumber \\
& = & W \log\left[ 1 + \frac{P_0}{P_N} \right]
- 2 W \gamma^2 P_0^2 \frac{P_0}{P_N} + {\cal O}\left(\gamma^4\right)
\ \ { \ } \ 
\label{eq:app_Cg_model}
\end{eqnarray}
which, as expected, is {\it smaller} that the actual channel capacity
(\ref{eq:app_Shannon}).
Note, that the difference between the exact channel capacity and the 
Gaussian estimate
\begin{eqnarray}
\delta C & \equiv & C - C_G = 
W \log\left[ 
1 + \frac{P_0}{P_N}
\left(1 - \frac{1}{\left(1 +  \gamma^2 P_0^2\right)^2}\right)
\right] \nonumber \\
& = & 2 W \gamma^2 P_0^2 \frac{P_0}{P_N} + {\cal O}\left(\gamma^4\right)
\ \ \ \ \label{eq:app_dC}
\end{eqnarray}
is not merely a constant scale factor, but a nontrivial function of the
signal to noise ratio, and the nonlinerity.

Even when the input distribution {\it is Gaussian}, like e.g. when
the phase $\phi$ depends on $x$ via the ``power'' $\left| x \right|^2$,
the Gaussian Estimate does not yield the exact result. The reason for
this behaviour is that the joint Gaussian distribution does not 
correctly reproduce the {\it conditional} distribution 
$p\left(y\left.\right|x\right)$.

For an essentially 
noinlinear  system (e.g. a fiber optics communication channel),
there is generally very little
{\it apriori} knowledge about the parametric dependence of the Channel Capacity
on the signal to noise ratio and other system parameters. In this case, 
the Gaussian Estimate for the channel capacity can be (should be?) viewed 
as a very unreliable method, as there is no way to separate it's
artefacts from the actual behaviour of the channel capacity.

\bibliographystyle{IEEE}

\end{document}